\documentclass[pdflatex,sn-mathphys-num,iicol]{sn-jnl}
\usepackage{graphicx}
\usepackage{multirow}
\usepackage{amsmath,amssymb,amsfonts}
\usepackage{amsthm}
\usepackage{mathrsfs}
\usepackage[title]{appendix}
\usepackage{xcolor}
\usepackage{textcomp}
\usepackage{manyfoot}
\usepackage{slashed}
\usepackage{booktabs}
\usepackage{caption} 
\usepackage{amsmath}
\usepackage{empheq}
\usepackage{algorithm}
\usepackage{algorithmicx}
\usepackage{algpseudocode}
\usepackage{listings}
\usepackage{tabularx}
\theoremstyle{thmstyleone}

\usepackage{tabularx,array}
\usepackage{etoolbox}
\AtBeginEnvironment{thebibliography}{\interlinepenalty=10000\relax}
\theoremstyle{thmstyletwo}

\usepackage{longtable}
\newcolumntype{Y}{>{\raggedright\arraybackslash}X}
\theoremstyle{thmstylethree}

\usepackage[noindentafter]{titlesec}

\begin{document}
\title[Dynamic Vacuum Energy in Weyl Integrable Spacetime Part I: Implications for Inflationary Cosmology]{Dynamic Vacuum Energy in Weyl Integrable Spacetime \\ Part I: Implications for Inflationary Cosmology}
\author*[1]{\fnm{Florian} \sur{Millo}}\email{florian.millo@cnrs.fr}
\affil*[1]{\orgdiv{Physics Department}, \orgname{Sorbonne Universit\'e (CNRS)}, \orgaddress{\street{4 place Jussieu}, \city{Paris}, \postcode{75005}, \country{France}}}

\abstract{We explore a dynamic vacuum energy scenario within the Weyl Integrable Spacetime (WIS) framework and provide the Friedmann--Lema\^itre (FL) cosmology equations under a metric endowed with a lapse function. We pay particular attention to the implications of inflationary dynamics. We show that the explicit vacuum decay source cancels in the FL acceleration equation, where the Hubble slow-roll parameter is controlled by the non-vacuum sector alone, while the dynamic vacuum defined in proper time $\Lambda(\tau)$ itself continues to enter the observables through the expansion rate. We identify the warm inflation (WI) dissipation coefficient with one half of the logarithmic decay rate of the vacuum energy, $\Upsilon=-\tfrac{1}{2} d\ln\Lambda/d\tau$, hence the phenomenological friction coefficient is tied to the vacuum sector rather than computed from explicit inflaton--bath microphysics. Notably, we show that the dissipation ratio $(Q)$ is not a free parameter: the cosmological constant hierarchy constrains it through $Q N_{e,\mathrm I}\simeq46$, where $N_{e,\mathrm I}$ is the total number of e-folds of expansion during vacuum decay. For a moderate pre-observable transient, this robustly gives $Q=\mathcal O(1)$. The graceful exit and the observable parameters are obtained and are in line with cosmological observations. These results indicate that our model is viable and provide a well-posed starting point for a subsequent numerical analysis.\\ \\ Published in  Eur. Phys. J. C 86, 1323 (2026) DOI:10.1140/epjc/s10052-026-16529-8}
\keywords{Dynamic Vacuum Energy, Dark Energy, Cosmological Constant Problem, Cosmic Inflation, Weyl Integrable Spacetime, Gravitation}

\maketitle
\section{Introduction}
\label{sec:intro}
\noindent The Big Bang theory traces the Universe from the initial singularity to
recombination, but cannot by itself resolve the flatness and horizon
problems. Both are addressed by a period of cosmic inflation
\cite{guth_inflationary_1981, linde_new_1982, albrecht_steinhardt_1982,
linde_hybrid_1994}, during which the potential energy of a slowly
rolling scalar field $\phi(t)$ sources a quasi de~Sitter phase of
accelerated expansion. The observed late time acceleration
\cite{riess_1998, perlmutter_1999} admits a formally identical
description, either as a bare cosmological constant $\Lambda$ or as a
dynamical quintessence field \cite{ratra_cosmological_1988,
wetterich_1988, caldwell_dave_steinhardt_1998}. Scenarios in which a
single canonical scalar drives both epochs are known as
quintessential inflation \cite{peebles_vilenkin_1999}, including warm
realisations \cite{dimopoulos_donaldson_wood_2019, rosa_ventura_2019}
and constructions based on emergent scale symmetry
\cite{rubio_wetterich_2017}. A geometrical way to introduce $\phi(t)$  into the Einstein-Hilbert (EH) action is through the Weyl Integrable Spacetime (WIS)
\cite{weyl_1918, dirac_1973, canuto_1977, novello_weyl_1983, salim_gravitational_1996, romero_general_2012,
paliathanasis_painleve_2022} in which $\phi(t)$ acts as the potential function of the non-metricity $w_\mu=\partial_\mu\phi$. WIS is
conformally equivalent to scalar--tensor theories \cite{brans_dicke_1961, horndeski_1974, fujii_maeda_2003, maeder_scale-invariant_2020} and to the
scale-invariant vacuum (SIV) hypothesis \cite{maeder_scale-invariant_2020,
Maeder2017, universe11020048}. This geometric setting therefore provides
a natural framework to examine the evolution of the vacuum
sector from the inflationary epoch to the present Universe. According to quantum field theory (QFT), the vacuum energy density during the inflationary period was vastly larger than its present value. This is known as the cosmological constant (CC) problem \cite{RevModPhys.61.1,
RevModPhys.75.559}. The main attempt to parametrize this problem is to assume a dynamic vacuum energy $\Lambda(t)$ encoding the evolution of the vacuum energy
from early to late times of the Universe. A review of dark energy dynamics is assembled in Refs.~\cite{PhysRevD.58.043506, COPELAND_2006}, and research
on dynamic vacuum energy has become increasingly active in recent years
\cite{PhysRevD.88.063531, doi:10.1142/9789811269776_0137, Gu_2025, desi_dr2_2025}. It is
therefore of high interest to implement a dynamic vacuum energy scenario within the (warm) inflationary paradigm.\\
\indent In this first part of the work, we develop a dynamic vacuum energy scenario under the WIS framework and provide the FL cosmology equations, paying particular attention to the implications for (warm) inflationary cosmology. The following results then emerge: (I) The vacuum decay contributions cancel in the FL acceleration equation, and as a consequence the first Hubble slow-roll parameter $(\epsilon_H)$ is fixed by the non-vacuum sector alone. (II) In the lapse gauge defined by the SIV hypothesis, $N(t)$ is related to the dynamical vacuum, while the gauge independent proper time logarithmic decay rate of that vacuum is phenomenologically identified with the WI dissipation ratio $Q=-d\ln \Lambda/(6Hd\tau)=\Upsilon/(3H)$ where $H$ is the Hubble parameter and $\Upsilon$ is the dissipation coefficient which reads as $\Upsilon=-\tfrac12 d\ln\Lambda/d\tau$. (III) The concave-shaped inflation potential is specified for a generic analysis that generates an attractor in which the spectral index $(n_s)$ depends only on the number of e-folds $N_e$ as $n_s\simeq1-3/(2N_e)$. (IV) The dissipation ratio is shown not to be a free parameter, the total decay of $\Lambda$ over the inflationary interval fixing the product $QN_{e,\mathrm I}$ and hence predicting $Q=\mathcal O(1)$. (V) The graceful exit is obtained quantitatively. Moreover, we interpret a correspondence between the integrable Weyl structure and holographic renormalization. In this interpretation the scalar beta function characterizes the field trajectory, while the dissipation ratio modifies the relation between the beta function and the Hubble slow-roll parameter. \\
\indent The remainder of this manuscript is organized as follows. In Sec.~\ref{EFEWeyl}, we introduce the WIS framework, obtain the modified EH action and impose the SIV hypothesis. In Sec.~\ref{FLCosmology}, the FL equations and the associated continuity laws are obtained. In Sec.~\ref{SecInflationaryC}, we derive the equation of motion for $\phi(t)$, define the dissipation ratio $Q$, specify a quasi-exponential inflaton potential, and compute the slow-roll parameters $(\epsilon_Q,\eta_Q)$, the spectral index $n_s$ and the tensor-to-scalar ratio $r_{\mathcal{T}}$, together with the graceful exit condition. In Sec.~\ref{sec:summary}, we discuss the physical content, the WIS-to-holographic correspondence and give an outlook. We then conclude in Sec.~\ref{conclusions}.

\section{Field Equations in Weyl Integrable Spacetime}\label{EFEWeyl}
Weyl Integrable Spacetime (WIS) provides the most natural way to include geometrically
a scalar degree of freedom into the gravitational
sector rather than adding it by hand as external matter
\cite{novello_weyl_1983, salim_gravitational_1996, romero_general_2012,
paliathanasis_painleve_2022}. Throughout this work we set $\hbar=c=M_{\rm Pl}=(8\pi G)^{-1/2}=1$, whence
$\kappa=8\pi G=1$. However, we retain $\kappa$ explicitly for dimensional readability. We endow the metric with a $(-+++)$ signature.
\subsection{Weyl Integrable Spacetime}
\label{subsec:weyl}
A Weyl manifold is a smooth manifold $\mathcal{M}$ equipped with $(g_{\mu\nu}, w_\mu)$ and a torsion-free affine connection $(\Gamma^\lambda_{\ \mu\nu})$ that is
not metric compatible \cite{weyl_1918}. Its non-metricity is controlled by the 1-form $w_\mu$,
\begin{align}
	\nabla_\lambda g_{\mu\nu} = -w_\lambda g_{\mu\nu} .
	\label{eq:nonmetricity}
\end{align}
The theory is integrable when $w_\mu = \partial_\mu \phi$ is exact, and $dw=0$, hence the associated Weyl scale curvature vanishes identically. Weyl integrability guarantees that the length of a vector parallel-transported around a closed loop is preserved \cite{Romero_2011}. The scalar $\phi$ is therefore not an independent matter field but the potential of the non-metricity itself. The geometry is invariant under the Weyl gauge transformations
\begin{align}
	g_{\mu\nu}\to e^{\Omega} g_{\mu\nu} ,
	\quad
	w_\mu \to w_\mu - \partial_\mu\Omega , \quad \phi\to\phi-\Omega,
	\label{eq:weylgauge}
\end{align}
for an arbitrary function $\Omega(x)$. These transformations generate the local Weyl group $\mathrm{Weyl}(\mathcal{M})\cong C^{\infty}(\mathcal{M},\mathbb{R}_{+})$, an infinite dimensional abelian gauge group which combines with the diffeomorphisms as the semidirect product $\mathrm{Diff}(\mathcal{M})\ltimes\mathrm{Weyl}(\mathcal{M})$. Weyl integrability implies trivial scale holonomy around a closed loop, which guarantees the absence of path dependent changes in vector lengths \cite{romero_general_2012}. The combination $\tilde g_{\mu\nu}=e^{\phi}g_{\mu\nu}$ is gauge invariant and defines the Weyl frame: with $w_\mu=\partial_\mu\phi$ the Weyl connection coincides with the Levi-Civita connection of $\tilde g_{\mu\nu}$. Contracting its Ricci tensor with the background metric, one obtains the 4-dimensional identities 
\begin{align}
	\tilde R &= R - 3\nabla^{2}\phi
	- \tfrac{3}{2}\nabla^\mu\phi\nabla_\mu\phi ,
	\label{eq:Rtilde}\\[2pt]
	g^{\mu\nu}\tilde\nabla_\mu\tilde\nabla_\nu\phi
	&= \nabla^{2}\phi + \nabla^\mu\phi\nabla_\mu\phi ,
	\label{eq:boxtilde}
\end{align}
where all quantities on the right are built from $g_{\mu\nu}$. Note that
$\tilde R = g^{\mu\nu}\tilde R_{\mu\nu}$ carries no conformal factor,
since the $e^{-\phi}$ factor of the curvature scalar of $\tilde g_{\mu\nu}$ is canceled by contracting with $g^{\mu\nu}$ rather than $\tilde g^{\mu\nu}$.
The identities themselves are standard textbook material \cite{Carroll_2019, faraoni_gunzig_nardone_1999, wald_1984}.
Throughout this work, we take as our starting point the modified Einstein--Hilbert (EH) action under WIS,
\begin{align}
	S = \int d^4x \sqrt{-g}
	\frac{1}{2\kappa}\Big[\tilde R
	+ 2\xi g^{\mu\nu}\tilde\nabla_\mu\tilde\nabla_\nu\phi
	- 2\Lambda(t) \Big] \nonumber\\[2pt]
	+ \int d^4x \sqrt{-g}
	\Big[ -V(\phi) + \mathcal{L}_{M} \Big] ,
	\label{eq:action_weyl}
\end{align}
with $\xi$ a non-minimal coupling constant, $\Lambda(t)$ the dynamic vacuum energy, $V(\phi)$ as the potential energy of the scalar degree of freedom and $\mathcal{L}_{M}$ as the Lagrangian of any matter fields. Two remarks on the matter sector constrain what $V(\phi)$ can be. First, matter couples minimally to $\tilde g_{\mu\nu}$ which when re-expressed in the background frame, gives the conformal weight
$e^{-\phi/2}$ carried by the stress-energy tensor [cf. Eq.~\eqref{eq:Tmunu}]. Second, a conformally invariant matter sector cannot source $\phi$, since radiation in
4-dimensions is traceless $(T^{\mu}{}_{\mu}=0)$, its action
is invariant under Weyl transformations [Eq.~\eqref{eq:weylgauge}] and hence the potential energy $V(\phi)$ must be introduced in the action. We now switch to the Einstein frame, where the conformal symmetry breaking factor is inherited.
\subsection{Einstein Field Equations and the Scale-Invariant Vacuum}
\label{subsec:einsteinframe}
\noindent Substituting Eqs.~\eqref{eq:Rtilde}--\eqref{eq:boxtilde} into
Eq.~\eqref{eq:action_weyl} and discarding the total derivative $(\nabla^2\phi)$, the action collapses to a single field scalar--tensor form in the background metric $g_{\mu\nu}$, obtaining the EH action,
\begin{align}
	S_{\rm EH} = \int d^4x \sqrt{-g}\bigg\{
	\frac{1}{2\kappa}\Big[R
	- \lambda\nabla^\mu\phi \nabla_\mu\phi
	- 2\Lambda(t)\Big] \nonumber\\[2pt]
	-V(\phi) + \mathcal{L}_{M} \bigg\} ,
	\qquad
	\lambda = \tfrac{3}{2} - 2\xi .
	\label{eq:action_einstein}
\end{align}
Here, the field $\phi$ is the potential of the integrable Weyl
non-metricity, $w_\mu=\partial_\mu\phi$, while its dynamics
are determined by the action, whose kinetic
normalization is $\lambda=3/2-2\xi$, with the contribution
$3/2$ inherited from Eq.~\eqref{eq:Rtilde}
\cite{romero_general_2012}. For $\lambda>0$ (a quintessence field \cite{ratra_cosmological_1988, wetterich_1988, caldwell_dave_steinhardt_1998}), corresponding to $\xi<3/4$, whereas
$\lambda<0$ (a phantom field \cite{paliathanasis_dynamical_2021,salim_gravitational_1996}) produces a wrong sign kinetic term. This can be seen if we perform a redefinition $\chi=\sqrt{\lambda/\kappa}\,\phi$ giving
the canonical positive kinetic energy term $-\tfrac12\nabla^\mu\chi\nabla_\mu\chi$. From the particle physics perspective, $\chi$ may be seen as an effective spin-0 degree of freedom whose association with conformal rescalings motivates a dilaton-like
particle, without identifying it with a known Standard
Model degree of freedom or a string dilaton.
Extremising Eq.~\eqref{eq:action_einstein} gives the Einstein field equations (EFE)
\begin{align}
	G_{\mu\nu}
	&+ \big[\Lambda(t) + \kappa V(\phi)\big] g_{\mu\nu} \nonumber\\
	&- \lambda\Big(\phi_{,\mu}\phi_{,\nu}
	- \tfrac{1}{2} g_{\mu\nu} \phi^{,\beta}\phi_{,\beta}\Big)
	= \kappa T^{M}_{\mu\nu},
	\label{eq:EFE}
\end{align}
where $T^{M}_{\mu\nu}$ is the perfect fluid stress-energy tensor defined as follows
\begin{align}
	&T^{M}_{\mu\nu} = e^{-\phi/2}
	\Big[ (\rho + p) u_\mu u_\nu + p g_{\mu\nu} \Big] ,
	\label{eq:Tmunu}\\
	&u^\mu u_\mu = -1, \nonumber
\end{align}
with $\{\rho, p, u\}$ being \{density, pressure, 4-velocity\} in the co-moving frame of reference with $u^\mu=\{N(t)^{-1}, 0,0,0\}$ described by the lapse function $N(t)$. We model the Universe using the spatially flat FLRW background metric,
\begin{align}
	g_{\mu\nu} = \textrm{diag}(-N(t)^2, a(t)^2,
	a(t)^2 r^2, a(t)^2 r^2\sin^2\theta) ,
	\label{eq:metric}
\end{align}
where $a(t)$ is the scale factor of the Universe.\\

\indent Throughout this work, we adopt the SIV hypothesis \cite{Maeder2017,
maeder_scale-invariant_2020, universe11020048, maeder_gueorguiev_2021}
\begin{align}
	\Lambda(t) = \Lambda(t_0) N(t)^2 ,
	\label{eq:scaleinvariant}
\end{align}
where $t_0$ denotes the present time. Eq.~\eqref{eq:scaleinvariant}
is used here as a lapse-gauge condition motivated by the
SIV hypothesis of
Refs.~\cite{Maeder2017,maeder_scale-invariant_2020,
universe11020048}. We emphasize that Eq.~\eqref{eq:scaleinvariant} is not a generally covariant identity for an arbitrary time coordinate. Indeed, under a time reparametrization, $t\rightarrow\bar t(t)$ the lapse transforms as $\bar N(\bar t)=N(t)dt/ d\bar t,$ while $\Lambda$ transforms as a scalar. Hence, Eq.~\eqref{eq:scaleinvariant} fixes the time parametrization rather than introducing an additional propagating degree of freedom. Once this gauge is chosen, $N(t)$ tracks the prescribed vacuum history according to $N(t)=\sqrt{\Lambda(t)/\Lambda(t_0)}$. All dynamical and observational statements below are ultimately written in terms of proper time, $d\tau=N(t)dt$, and the lapse disappears from the physical evolution equations except through the chosen function $\Lambda(\tau)$. More specifically, the construction given in Eq.~\eqref{eq:scaleinvariant} specifies the time parametrization of a given vacuum history, without introducing an additional propagating degree of freedom. The SIV hypothesis does not itself determine the physical evolution of the vacuum energy. That evolution is specified separately in proper time and must satisfy the total energy balance required by the Bianchi identity. In particular, neither vacuum decay nor physical dissipation can be inferred from the lapse choice alone. The authors of Ref.~\cite{maeder_scale-invariant_2020} present the SIV hypothesis as a minimal framework based on the scale invariance of empty space, reporting agreement with observational tests and identifying further testable predictions.\\
\indent Before proceeding further, we fix the notation for a symbol that recurs in three roles. Throughout this manuscript, $N(t)$ denotes the lapse function of Eq.~\eqref{eq:metric}, $N_e$ denotes the number of e-folds before the end of inflation, and $N_c$ denotes the e-fold at which the Lambert branch point [Eq.~\eqref{GammaExplicit}] is reached. These three notations are never used interchangeably.\\
\indent A few consequences of EFEs under WIS [Eq.~\eqref{eq:EFE}] are worth discussing. Eq.~\eqref{eq:EFE} depends on $\Lambda$ and $V$ through the
combination
\begin{align}
	\Lambda_{\rm eff}(t,\phi) = \Lambda(t) + \kappa V(\phi) .
	\label{eq:Lambdaeff}
\end{align}
The two terms are algebraically degenerate at the level of the
field equations and can be distinguished only by their evolution: both $V(\phi(t))$ and $\Lambda(t)$ are time-dependent, but the former is implicit whereas the latter is explicit. This is the case where a vacuum energy depending
on coordinate time is not covariant. A fully covariant formulation is recovered by promoting $\Lambda \to \Lambda(\phi)$, in which case [Eq.~\eqref{eq:Lambdaeff}] shows that the dynamic vacuum energy can be
reabsorbed into the scalar potential. However, we keep the two terms separate because they play distinct physical roles in our model: $V$ drives inflation, while $\Lambda$ now supplies a dissipation channel and most importantly drives the expansion of the Universe. Applying the covariant derivative to Eq.~\eqref{eq:EFE} and using $\nabla^\mu G_{\mu\nu}=0$ gives the running vacuum continuity law
\begin{align}
	\dot\rho_{\rm eff} + 3HN\big(\rho_{\rm eff} + p_{\rm eff}\big)
	= -\dot\rho_\Lambda ,\; \rho_\Lambda = \Lambda/\kappa. 
    \label{eq:exchange}
\end{align}
Eq.~\eqref{eq:exchange} is forced by $\nabla^\mu G_{\mu\nu}=0$, once $\Lambda$ is allowed to vary \cite{PhysRevD.88.063531}. An important hint lies within Eq.~\eqref{eq:exchange}: in a strictly vacuum dominated
background, $\rho_{\rm eff}=0$ forces $\dot\Lambda = 0$; a decaying
vacuum energy is therefore incompatible with exact vacuum domination and requires a non-vacuum sector into which the vacuum energy can drain. We discuss the consequences of dynamic vacuum energy under WIS for the FL cosmology equations.
\section{Friedmann--Lema\^itre Cosmology Equations} \label{FLCosmology}
In this section, we provide the FL cosmology equations within the WIS framework. We explore a homogeneous and isotropic Universe by adopting the FLRW metric [Eq.~\eqref{eq:metric}]. 
For brevity, we omit writing the non-zero Christoffel terms, the explicit form of the Riemann tensor and Ricci curvature tensor. Nonetheless, for reasons that will become clear later, we show only the Ricci scalar
\begin{align}
	R &= -\frac{6}{N^2}\left( \frac{\dot a}{a} \frac{\dot N}{N} - \left(\frac{\dot a}{a}\right)^2 -\frac{\ddot a}{a} \right).  \label{ricciscalar}
\end{align}
The FL and continuity equations deduced from the WIS framework are
\begin{align}
 &H^2 = \frac{\kappa}{3}\rho_{\rm eff} + \frac{\Lambda}{3},
      \label{FL1} \\[4pt]
 &\frac{\dot H}{N} = -\frac{\kappa}{2}\left(\rho_{\rm eff}+p_{\rm eff}\right)
      = -\frac{2 \kappa}{3} e^{-\phi/2}\rho_{\rm rad}
      -\frac{\lambda \dot\phi^{2}}{2N^{2}},
      \label{FL2} \\[4pt]
 & \dot\rho_{\rm rad} + 4HN\rho_{\rm rad}
      = \tfrac{1}{2}\dot\phi\rho_{\rm rad} -e^{\phi/2}\left(\frac{\lambda}{\kappa}\frac{\dot N}{N}\frac{\dot\phi^{2}}{N^{2}}
      + \dot \rho_{\Lambda}\right),
      \label{MRContEquations}
\end{align}
where $H=H(a(t),N(t))=\dot a/(Na)$ is the Hubble parameter and
\begin{align}
\rho_{\rm eff} &= e^{-\phi/2}\rho_{\rm rad}
  + \frac{\lambda\dot\phi^{2}}{2\kappa N^{2}} + V(\phi) , \nonumber\\
p_{\rm eff} &= \tfrac{1}{3}e^{-\phi/2}\rho_{\rm rad}
  + \frac{\lambda\dot\phi^{2}}{2\kappa N^{2}} - V(\phi),
  \label{eq:effdens}
\end{align}
denote the non-vacuum content. We stress the following distinction: $\rho_{\rm rad}$ is the radiation density of the bath, whereas $\rho_{\rm eff}$ is the full non-vacuum density including the kinetic and potential contributions of the scalar. We restrict our study to the radiation content of the early Universe, with pressure density $p=w\rho$ and equation of state $w=1/3$. 

The standard FL equations can be retrieved when $\{\phi\to0, \Lambda(t)\to\Lambda\}$ are applied to Eq.~\eqref{FL1} and Eq.~\eqref{FL2}. The acceleration equation [Eq.~\eqref{FL2}] contains neither $\dot\Lambda$ nor $\dot N$. Differentiating Eq.~\eqref{FL1} and inserting Eq.~\eqref{eq:exchange}, the explicit vacuum decay source $\dot\rho_\Lambda$ cancels, leaving $\dot H/N=-\kappa(\rho_{\rm eff}+p_{\rm eff})/2$. What cancels is the explicit source $\dot\rho_\Lambda$, while the value of $\Lambda$ itself survives in the dimensionless slow-roll parameters through $H^{2}$, as Eq.~\eqref{eq:exitcondition} below makes explicit.
Consequently, the first Hubble slow-roll parameter $(\epsilon_{H})$ may be controlled solely by the non-vacuum content. Vacuum decay redistributes
energy between sectors without directly decelerating the expansion, hence a dissipation parameter $(Q)$ might constitute a degree of freedom independent of $\epsilon_{H}$. A few points are worth highlighting.

First, Eq.~\eqref{ricciscalar} displays the usual lapse dependence of the Ricci scalar when an arbitrary time coordinate is used. In the lapse gauge of Eq.~\eqref{eq:scaleinvariant}, this coordinate dependence can be expressed in terms of the prescribed
vacuum history through $N=\sqrt{\Lambda/\Lambda(t_0)}$. This does not make the lapse itself a physical observable. The physical influence of the dynamical vacuum is instead encoded in the EFEs and in the evolution of $\{H,\rho_{\rm eff},\Lambda\}$.\\

\indent Second, the late time behavior ($t\to t_{0}$, now) of the FL expansion equation [Eq.~\eqref{FL1}] admits a solution in which the scalar
field asymptotes to a small positive constant, the precise value depending
on $V(\phi)$ and on the present day parameters $\{H(t_{0}),\rho_{\rm eff}(t_{0}),\Lambda(t_{0})\}$. The full numerical analysis is deferred to the second part of this work. Eqs.~\eqref{FL1} and \eqref{FL2} demonstrate that a scalar field once introduced in the early Universe remains a permanent feature of the cosmic expansion, contributing to the background dynamics indefinitely.\\
\indent Third, Eqs.~\eqref{FL1}--\eqref{MRContEquations}
admit a de Sitter solution in the vacuum limit $\rho_{\rm eff}=0$ and $\phi=0$, provided that $\Lambda$ is constant. The choice of the vacuum law, which is later introduced, should be consistent with an asymptotic de Sitter limit, as $\Lambda(\tau)$ approaches the positive constant $\Lambda_\infty=\lim_{\tau\to\infty}\Lambda(\tau)$.\\
\indent Fourth, it is of interest how the cosmic acceleration/deceleration varies in our model. To this end, the deceleration parameter $(q)$ is of high importance. Using Eq.~\eqref{FL1} and \eqref{FL2}
\begin{align}
q&=-\frac{a a^{\prime \prime}}{a^{\prime 2}}=-1-\frac{H^{\prime}}{H^2}\nonumber \\
&=-1+\frac{\kappa}{2 H^2}\left(\chi^{\prime 2}+\frac{4}{3} \varrho_R\right)=-1+\epsilon_H
\end{align}
where $'=d/d\tau$. This exact result gives $q=-1$ for the de Sitter point, $q<0$ during accelerated expansion, and $q=0$ at the end of acceleration. A radiation-dominated epoch, if reached dynamically, has $q \rightarrow 1$. In arbitrary coordinate time the equivalent expression is $q=-a \ddot{a} / \dot{a}^2+a \dot{N} /(N \dot{a})$. A complete curve $q\left(N_e\right)$ requires the numerical background solution and will not be inferred solely from the limiting values. \\
\indent To gain more insight into the scalar field dynamics of the early Universe, we examine the warm inflationary paradigm --- where a dissipation mechanism defined through the decay rate of the vacuum energy is identified --- in the following section.
\section{Implications for Inflationary Cosmology}\label{SecInflationaryC}
\noindent In this section, we embed the WIS scalar sector in an effective warm inflation (WI) scenario, in which the canonically normalized inflaton field evolves in the presence of a dissipative interaction with a radiation bath. The underlying derivation is straightforward. Variation of Eq.~\eqref{eq:action_einstein} with respect to $\phi$ yields the standard cold inflation equation of motion. Then, introducing the canonically normalized field $\chi=\sqrt{\lambda/\kappa} \phi$ and expressing the dynamics in proper time, we obtain $\chi^{\prime \prime}+3 H \chi^{\prime}+d V/d\chi=0$. Finally, we embed this scalar sector in the WI effective dynamics \cite{berera_moss_ramos_2009, kamali_2023} by introducing a dissipation ratio $(Q)$ as follows, 
\begin{align}
	\chi'' + 3H\left(1+Q\right)\chi' + \frac{d V}{d\chi} = 0
	\label{eq:KGproper}
\end{align}
where $Q=\Upsilon/(3H)$ and $\Upsilon$ is the dissipation coefficient. As in standard WI treatments, the dissipative contribution is associated with the interaction of the inflaton with a radiation bath and does not follow from the isolated scalar sector of Eq.~\eqref{eq:action_einstein}. Whereas $\Upsilon$ is conventionally computed from inflaton--bath interactions and inserted into Eq.~\eqref{eq:KGproper}, here we propose instead the identification
\begin{align}
	&Q = -\frac{1}{6HN}\frac{\dot\Lambda}{\Lambda} = \frac{\Upsilon}{3H} ,
	\label{q0dissipation}\\
	&\Upsilon = -\frac{1}{2N}\frac{\dot\Lambda}{\Lambda}
	= -\frac{1}{2}\frac{d \ln\Lambda}{d \tau} ,
	\label{qdissipation}
\end{align}
where the dissipation coefficient is equal to one half of the logarithmic decay rate of the vacuum energy. The coordinate time representation in Eq.~\eqref{q0dissipation} contains the lapse explicitly; however, the combination $\dot\Lambda/N=d\Lambda/d\tau$ is invariant under a reparametrization of the time coordinate. Therefore, the proper time expressions $\{Q(\tau),\Upsilon(\tau)\}$ are the gauge independent quantities entering the effective WI description.\\
\indent We emphasize that Eq.~\eqref{qdissipation} is an identification rather than a derivation from the action of Sec.~\ref{EFEWeyl}. In standard WI treatments, by contrast, $\Upsilon$ is obtained or parametrized from the underlying inflaton--bath interactions \cite{berera_warm_1995, berera_fang_1995, berera_moss_ramos_2009, bastero_gil_2021}. Nevertheless, the identification made in Eq.~\eqref{qdissipation} is dynamically consistent with the energy balance of the effective WI system. The inflaton loses energy according to $\rho_{\chi}'+3H(\rho_{\chi}+p_{\chi})=-\Upsilon\chi'^{2}$, while the corresponding term enters the radiation sector with the opposite sign. Because the radiation density appears in EFEs through the conformally weighted combination $e^{-\phi/2}\rho_{\rm rad}$, Eq.~\eqref{MRContEquations} contains the corresponding factor $e^{\phi/2}$ in its source terms. Adding the inflaton, radiation, and vacuum equations then yields the total conservation law. Under the SIV framework [Eq.~\eqref{eq:scaleinvariant}] one has $\Upsilon=-\dot N/N^{2}$, where $\Upsilon>0$ requires $\dot N<0$. The dissipation carries the correct sign automatically, without an additional assumption. \\
\indent The main hypothesis of this work is Eq.~\eqref{qdissipation}: the dissipation coefficient of a WI scenario is identified, in proper time, with one half of the logarithmic decay rate of the vacuum energy. Whereas in the standard construction $\Upsilon=\Upsilon(T)$ -- where $T$ is the temperature of the radiation bath during inflation -- encodes the microphysics of inflaton--bath interactions, in the present effective description we identify $\Upsilon$ with a quantity already determined by the dynamical vacuum sector [Eqs.~\eqref{qdissipation} and \eqref{q0dissipation}]. Thus, the dissipation coefficient is not derived from the isolated WIS action, but its phenomenological value is tied to the vacuum dynamics rather than introduced as an independent function. We stress the distinction from the WI formulated within modified gravity, including the recent treatment in Weyl geometric gravity \cite{huang_harko_2026}, where $\Upsilon\propto T$ is postulated alongside the modified gravitational sector. In our model, $\Upsilon$ is likewise adopted rather than computed from first principles, but it is identified with a quantity already present in the gravitational sector rather than introduced independently of it [Eq.~\eqref{qdissipation}]. \\ 
\indent We now specify the dynamic vacuum energy law adopted throughout this work. QFT indicates that during inflation $\Lambda_{\textrm{I}}$ exceeded its present value $\Lambda(t_0)$ by some $120$ orders of magnitude \cite{RevModPhys.61.1}, decaying similarly to a ``cooling law''. The precise exponent, quoted variously as $10^{120}$, $10^{122}$ or $10^{123}$, depends on whether reduced or non-reduced Planck units are used and on whether one compares $\Lambda$ or $\rho_\Lambda$. We retain $\Lambda_{\textrm{I}}/\Lambda_\infty\sim10^{120}$ and show in Sec.~\ref{subsec:Qfixed} that our results depend on this choice only logarithmically. To this end, we assume the following dynamic vacuum law,
\begin{align}
	\Lambda(\tau) = \Lambda_{\infty}
	+ \big(\Lambda_{\textrm{I}} - \Lambda_{\infty}\big)e^{-\gamma \tau} ,
	\quad \gamma = \textrm{const} ,
	\label{eq:Lambdaansatz}
\end{align}
where $\Lambda_\infty=\lim_{\tau\to\infty}\Lambda(\tau)$ denotes the asymptotic value of the vacuum energy. From Eq.~\eqref{eq:Lambdaansatz}, we observe that it is calibrated at $\tau=0$ where $\Lambda_\infty$ denotes the asymptotic value. We emphasize that Eq.~\eqref{eq:Lambdaansatz} is a phenomenological assumption based on the ``cooling" law dynamics (a first-order relaxation equation of type $\Lambda^{\prime}=-\gamma\left(\Lambda-\Lambda_{\infty}\right),\,\gamma>0$). The inflationary calibration gives $\gamma\sim H_{\rm I}\gg H_0$, thus by the present day value $\gamma\tau_0\gg1$. Moreover, $\Lambda(t_0)-\Lambda_\infty=(\Lambda_{\mathrm I}-\Lambda_\infty)e^{-\gamma\tau_0} \ll \Lambda_\infty$ and $\Lambda(t_0)$ becomes observationally indistinguishable from $\Lambda_\infty$. (We keep both symbols in order to distinguish the finite time from its asymptotic value.) At $\tau\to0^+$, $\Lambda(\tau)$ goes towards $\Lambda_{\textrm{I}}$ value. Eq.~\eqref{GammaExplicit} below then determines the $\gamma$-value from $(H_\ast,Q_\ast)$ at horizon crossing. Inserting Eq.~\eqref{eq:Lambdaansatz} into
Eq.~\eqref{q0dissipation} and solving for $\gamma$, we obtain
\begin{align}
	\gamma = 6HQ - \frac{1}{\tau}W_{0}\left[6HQ\tau e^{6HQ\tau}
	\frac{\Lambda_{\infty}}{\Lambda_{\infty}-\Lambda_{\textrm{I}}}\right],
	\label{GammaExplicit}
\end{align}
where $W_{0}$ denotes the principal branch of the Lambert function \cite{corless_1996}. Two remarks on the domain of Eq.~\eqref{GammaExplicit} are necessary: 1) since $\Lambda_{\infty}<\Lambda_{\textrm{I}}$, the argument of $W_{0}$ is negative; 2) the real branches exist for arguments larger than $-1/e$. The bound is saturated at a time $\tau_c$ given by
\begin{equation}
\begin{split}
    6HQ\tau_{c}
    &\simeq \ln(\Lambda_{\textrm{I}}/\Lambda_{\infty}) - 1 ,\\[2pt]
    N_{c} = H\tau_{c}
    &\simeq \frac{1}{6Q}\ln(\Lambda_{\textrm{I}}/\Lambda_{\infty}) .
\end{split}
\label{eq:branchpoint}
\end{equation}
Beyond $\tau_c$, the constant $\gamma$ inversion admits no real solution. We therefore treat Eq.~\eqref{GammaExplicit} as an early time
calibration relation, evaluated once at horizon crossing to fix $\gamma$
in terms of $(H_{\ast},Q_{\ast})$, rather than as a time evolution. In the early time limit $\tau\to0^{+}$, using $W_{0}(z)\simeq z$, one finds
\begin{align}
	\gamma \to 6H_{\textrm{I}}Q
	\frac{\Lambda_{\textrm{I}}}{\Lambda_{\textrm{I}}-\Lambda_{\infty}}
	\simeq 6H_{\textrm{I}}Q.
	\label{eq:gammaearly}
\end{align}
Since the predicted dissipation ratio is $Q=\mathcal O(1)$, Eq.~\eqref{eq:gammaearly} implies $\gamma \sim H_{\rm I}\gg H_0$. Thus $\gamma$ is not a late time slow decay scale, but it characterizes
the rapid primordial relaxation of the vacuum sector. The observed
late time constancy of $\Lambda$ follows because this transient has become exponentially suppressed long before the present time. Using Eqs.~\eqref{qdissipation} and \eqref{eq:Lambdaansatz}, the
dissipation coefficient can be written explicitly as
\begin{align}
    \Upsilon(\tau)
    &=
    \frac{\gamma}{2}
    \frac{(\Lambda_{\mathrm I}-\Lambda_\infty)e^{-\gamma\tau}}
    {\Lambda(\tau)}
    =
    \frac{\gamma}{2}
    \left(1-\frac{\Lambda_\infty}{\Lambda(\tau)}\right).
    \label{eq:UpsilonExact}
\end{align}
By virtue of Eq.~\eqref{eq:UpsilonExact}, whenever $\Lambda\gg\Lambda_\infty$, and in particular over the
observable slow-roll window considered below, $\Upsilon\simeq\gamma/2$ is approximately constant. Moreover, $\Upsilon$ carries no explicit temperature dependence, thus the temperature dependence parameter
$c=T\Upsilon_{,T}/\Upsilon$ \cite{graham_moss_2009} vanishes in the effective WI description. The temperature-driven growing mode associated with $c>0$ is therefore absent at leading order in our model.
\subsection{The dissipation ratio is constrained by the dynamical vacuum}
\label{subsec:Qfixed}
Writing Eq.~\eqref{q0dissipation} in proper time,
$d\tau=Ndt$, and using
$dN_e/d\tau=-H$, we obtain
\begin{align}
    \frac{d\ln\Lambda}{dN_e}
    =6Q(N_e).
    \label{eq:QNe}
\end{align}
Hence, over the complete vacuum decay interval,
\begin{align}
    \ln\left(\frac{\Lambda_{\rm I}}{\Lambda_{\rm end}}\right)
    =6\int_{0}^{N_{e,\rm I}}Q(N_e) dN_e.
    \label{eq:Qintegral}
\end{align}
For an approximately constant dissipation ratio over the interval,
we obtain the reduced form
\begin{align}
    \Lambda(N_e)=
    \Lambda_{\rm I}
    e^{-6Q(N_{e,\rm I}-N_e)}, \label{eq:LambdaNe}
\end{align}
where $N_{e,\mathrm{I}}$ denotes the total number of e-folds remaining
at the beginning of the vacuum decay interval considered here, such that
$\Lambda(N_{e,\mathrm{I}})=\Lambda_{\mathrm{I}}$, while
$N_e\rightarrow0$ at the end of inflation. The total decay of the vacuum energy across the vacuum decay interval is therefore not at our disposal but is fixed by the cosmological constant hierarchy. Under the working assumption that the vacuum decay is completed by the end of inflation, such that $\Lambda_{\rm end}\simeq\Lambda_\infty$, requiring $\Lambda$ to fall from $\Lambda_{\textrm{I}}$ to $\Lambda_{\rm end}$ gives
\begin{align}
    Q = \frac{1}{6N_{e,\textrm{I}}}\ln\left(\frac{\Lambda_{\textrm{I}}}
    {\Lambda_{\textrm{end}}}\right)\simeq\frac{1}{6N_{e,\textrm{I}}}
    \ln\left(\frac{\Lambda_{\textrm{I}}}{\Lambda_{\infty}}
    \right)\simeq\frac{46}{N_{e,\textrm{I}}},
    \label{eq:Qpred}
\end{align}
where $\Lambda_{\textrm{I}}/\Lambda_{\infty}\sim10^{120}$ is used. Eqs.~\eqref{eq:Qpred} and \eqref{eq:branchpoint} agree at leading
logarithmic order but represent distinct conditions. The former fixes
the integrated vacuum decay, while the latter marks the endpoint of
the real Lambert inversion. Strictly speaking, Eq.~\eqref{eq:Qpred} constrains the e-fold averaged
dissipation ratio over the complete vacuum decay interval. We identify this average with the local observable value $Q_\ast$, since
$\Lambda\gg\Lambda_\infty$ and $H$ varies only at slow-roll order during
the observable window. The accuracy of $Q_\ast\simeq Q$ will be tested
by the full numerical background evolution. \\
\indent Our model is self-consistent on the hyperbola
$QN_{e,\mathrm{I}}\simeq46$, for which the Lambert branch point [$W_0$ function of Eq.~\eqref{GammaExplicit}] lies
near the end of the assumed primordial vacuum decay interval, to the
logarithmic accuracy used in Eq.~\eqref{eq:Qpred}. From  Eqs.~\eqref{eq:scaleinvariant} and \eqref{eq:QNe}, the lapse evolves as $N\left(N_e\right)=N_{\mathrm{I}} e^{-3 Q\left(N_{e, \mathrm{I}}-N_e\right)}$. Hence, at the end of inflation the lapse is $N_{\mathrm{end}}=N_{\mathrm{I}} e^{-3 Q N_{e, \mathrm{I}}}$. For the representative values $\{N_{e,\mathrm{I}},Q\}=\{72,0.64\}$, one obtains
$N_{\rm end}\sim10^{60}\times10^{-60}\simeq1$ [See Eq.~\eqref{eq:NeDecomposition} for the definition of $N_{e,\mathrm{I}}$]. 
Within the lapse gauge of Eq.~\eqref{eq:scaleinvariant}, this simply
reflects the assumed completion of the vacuum decay,
$\Lambda_{\rm end}\simeq\Lambda_\infty\simeq\Lambda(t_0)$, together
with the normalization $N(t_0)=1$. This is not an additional physical
prediction. We distinguish the beginning of the full vacuum decay interval, denoted by $N_{e,\mathrm{I}}$, from the observable horizon crossing epoch $N_{e,\ast}$. In general, $N_{e,\mathrm{I}}\geq N_{e,\ast}$. The hierarchy relation given in Eq.~\eqref{eq:Qpred} constrains the total decay interval $N_{e,\mathrm{I}}$, whereas the inflationary observables are evaluated at $N_{e,\ast}$. We therefore allow for an initial vacuum decay transient during which $\Lambda$ decreases rapidly from $\Lambda_{\rm I}$. This early stage is not assumed to satisfy the WI slow-roll approximation employed below. Eq.~\eqref{eq:Qintegral} counts e-folds of expansion and is therefore
unaffected. The observable slow-roll regime is reached once
\begin{align}
    \frac{\Lambda_\ast}{\kappa V_\ast} \ll 1 ,
    \label{eq:LambdaSubdominant}
\end{align}
where the subscript $\ast$ denotes evaluation at the horizon crossing. Accordingly, the total interval may be decomposed as
\begin{align}
    N_{e,\mathrm{I}} = N_{e,\ast} + \Delta N_{\rm pre},
    \label{eq:NeDecomposition}
\end{align}
where $\Delta N_{\rm pre}$ denotes the duration of the initial
vacuum decay transient. \\
\indent Four consequences follow: (I) the hierarchy relation fixes
$Q\simeq46/N_{e,\mathrm{I}}$, placing the dissipation ratio naturally at
$\mathcal{O}(1)$ for a total vacuum decay interval of order $50$--$75$ e-folds; a moderate pre-observable transient modifies
the precise value of $Q$ but not its order of magnitude. (II) Strongly dissipative values are incompatible with a $\mathcal{O}(50)$-e-fold decay interval for the assumed hierarchy: for $N_{e,\mathrm{I}}=50$, $Q=10$ and $Q=10^2$ would require $\Lambda_{\mathrm{I}}/\Lambda_\infty$ of order $10^{1303}$ and $10^{13029}$, respectively. (III) The prediction is robust against the convention ambiguity noted above: across $\Lambda_{\textrm{I}}/\Lambda_{\infty}=10^{120}$--$10^{123}$ the product $QN_{e,\mathrm{I}}$ shifts only from $46.1$ to $47.2$, that is by under $3\%$. For the representative ACT value $Q_\ast\simeq0.64$, the vacuum evolves at $|d\ln\Lambda/dN_e|=6Q_\ast\simeq3.84$ per e-fold, corresponding to approximately $1.67$ decades per e-fold. The assumed $120$-decade hierarchy is therefore traversed over $N_{e,\rm I}\simeq72$ e-folds, consistently with
$N_{e,\rm I}=N_{e,\ast}+\Delta N_{\rm pre}
\simeq58+13.8$ [see Table~\ref{TableEpsvEtav}]. (IV) The Lambert branch point lies [Eq.~\eqref{GammaExplicit}] parametrically near the end of the assumed vacuum decay interval. The working identification of this event with the physical end of inflation must be checked independently through the condition $\epsilon_H=1$.
\subsection{Inflaton potential and slow-roll trajectory}
\label{subsec:potential}
\noindent The potential energy of $\phi$ is $V(\phi)$ and must be specified in the action [Eq.~\eqref{eq:action_einstein}]. To qualitatively demonstrate the robustness of our model, we adopt the one-parameter plateau potential
\begin{align}
    V(\chi) =
    V_{0}\left[1-e^{-\alpha\chi}\right]^{2},
    \qquad
    \chi=\sqrt{\lambda/\kappa}\phi .
    \label{potentialscalar}
\end{align}
This potential energy [Eq.~\eqref{potentialscalar}] belongs to the quasi-exponential class of single field inflationary
models \cite{MARTIN2024101653}. This class of potentials may arise in $f(R,T)$
gravity~\cite{ahmed_investigating_2024} and in Cartan $f(R)$
gravity~\cite{inagaki_cartan_2022}. On the large field plateau relevant for inflation, Eq.~\eqref{potentialscalar} is concave, while it possesses a stable minimum at $\chi=0$. Concave potentials on the large field inflationary plateau are favored by current cosmological data \cite{ellis2026inflation2025}. We emphasize that plateau and $\alpha$-attractor potentials have already been studied within WI \cite{kamali_2023, benetti_ramos_2017, berghaus_2020, chakraborty_das_2025, saha_nandy_2025}. Our novelty lies not in the choice of $V(\chi)$ but in the identification of the WI dissipation coefficient $\Upsilon$ with the logarithmic decay rate of the dynamical vacuum [Eq.~\eqref{qdissipation}], and in the resulting constraint on $Q$ [Eq.~\eqref{eq:Qpred}]. Moreover, the exponential dependence entering
Eq.~\eqref{potentialscalar} is motivated by Noether symmetry analyses of
WIS field equations, where exponential scalar
potentials lead to Liouville integrable structures
\cite{paliathanasis_noether_2022}. The Noether symmetry result motivates the exponential dependence, while the plateau structure adopted here follows from the squared quasi-exponential form of Eq.~\eqref{potentialscalar}. The WIS kinetic normalization $\lambda$ enters through the canonical relation $\chi=\sqrt{\lambda/\kappa} \phi$. The exponent and constant terms of $V(\chi)$, $\{\alpha,V_0\}$ in Eq.~\eqref{potentialscalar}, play the following roles: the plateau
$V\to V_{0}$ at $\alpha\chi\gg1$ sustains the slow-roll dynamics, while the minimum at $\chi=0$ terminates it. Setting $y=e^{-\alpha\chi}\ll1$,  the slow-roll trajectory can be obtained analytically. Using Eq.~\eqref{eq:KGproper} in the slow-roll regime together with $dN_e=-H d\tau$ and using Eq.~\eqref{FL1} as a constraint, one obtains
\begin{align}
    \frac{d\chi}{dN_e}
    \simeq
    \frac{V_{,\chi}}
    {(1+Q)\left(\kappa\rho_{\rm eff}+\Lambda\right)} .
\end{align}
During the observable slow-roll regime, the radiation and kinetic
contributions to $\rho_{\rm eff}$ are of order
$\mathcal{O}\left(\epsilon_V/(1+Q)^2\right)$. We further impose the
subdominance condition of Eq.~\eqref{eq:LambdaSubdominant},
$\Lambda/(\kappa V)\ll1$, after the initial vacuum decay transient
discussed in Sec.~\ref{subsec:Qfixed}. Hence,
$\kappa\rho_{\rm eff}+\Lambda\simeq\kappa V$, and therefore
\begin{align}
    \frac{d\chi}{dN_e}
    \simeq
    \frac{1}{1+Q}\frac{1}{\kappa V}\frac{dV}{d\chi}.
\end{align}
Using $\kappa=1$ in reduced Planck units and
\begin{align}
    \frac{V_{,\chi}}{V}
    =
    \frac{2\alpha y}{1-y}
    \simeq 2\alpha y ,
\end{align}
on the inflationary plateau, the slow-roll trajectory becomes
\begin{align}
    \frac{d\chi}{dN_e}
    &\simeq\frac{2\alpha y}{1+Q},
    \qquad
    \frac{d(1/y)}{dN_e}
    \simeq\frac{2\alpha^2}{1+Q}.
\end{align} 
\noindent Neglecting the integration constant in the large-$N_e$ plateau regime, where $2\alpha^2N_e/(1+Q)\gg1$, one obtains
\begin{align}
    y(N_{e}) \simeq
    \frac{1+Q}{2\alpha^{2}N_{e}}, \quad
    \chi(N_{e}) \simeq
    \frac{1}{\alpha}
    \ln\!\left(
    \frac{2\alpha^{2}N_{e}}{1+Q}
    \right).
    \label{eq:trajectory}
\end{align}
The asymptotic form [Eq.~\eqref{eq:trajectory}] demands that
$1+Q\ll2\alpha^{2}N_e$, which at the predicted $Q\simeq1$
reduces to $1\ll\alpha^{2}N_e$ [Eq.~\eqref{eq:Qpred}].
For $N_e\approx50$, this condition is comfortably satisfied for
$\alpha\gtrsim\mathcal O(1)$ in reduced Planck units. We retain $\alpha=\mathcal{O}(10)$ in the numerical estimates below, hence the $\epsilon_{Q}$ correction to the tilt remains at the $10^{-5}$ level. At $\alpha=1$ it grows to about $2\times10^{-3}$ and displaces the spectral index $n_{s}$ by a comparable amount.
\subsection{Slow-roll parameters and observables}
\label{subsec:observables}
In the WI regime, the potential slow-roll
parameters $\{\epsilon_V,\eta_V\}$ are suppressed by $(1+Q)$ \cite{berera_warm_1995, bastero_gil_2021},
\begin{equation}
\begin{split}
    \epsilon_{Q} &= \frac{\epsilon_V}{1+Q}
	= \frac{1}{2(1+Q)}\left(\frac{V_{,\chi}}{V}\right)^{2} ,\\[2pt]
	\eta_{Q} &= \frac{\eta_V}{1+Q}
	= \frac{1}{1+Q}\frac{V_{,\chi\chi}}{V} .
\end{split}
	\label{epsiloncond}
\end{equation}
Evaluating Eq.~\eqref{epsiloncond} on the slow-roll trajectory [Eq.~\eqref{eq:trajectory}] gives the following compact result,
\begin{align}
	\epsilon_{Q} \simeq \frac{1+Q}{2\alpha^{2}N_{e}^{2}} ,
	\qquad
	\eta_{Q} \simeq  -\frac{1}{N_{e}} .
	\label{eq:sreval}
\end{align}
At leading order in the large-field plateau expansion,
$\eta_Q$ depends on neither $\alpha$ nor $Q$. The cancellation follows
because $ V_{,\chi\chi}/V \simeq -2\alpha^2 y ,$ while the slow-roll trajectory gives $y\simeq(1+Q)/(2\alpha^2N_e)$. Consequently,
$\eta_Q\simeq-1/N_e$ independently of $\alpha$ and $Q$ at this order.
At fixed $N_e$, increasing $\alpha$ shifts the trajectory toward smaller $y$, farther out on the plateau, compensating for the $\alpha^2$ dependence of the potential curvature. Conversely, increasing $Q$ shifts the trajectory toward larger $y$, closer to the edge of the plateau and hence toward a larger curvature magnitude. This compensates the explicit
$(1+Q)^{-1}$ suppression in $\eta_Q$. These effects cancel at leading
order, thus yielding $\eta_Q\simeq-1/N_e$.
The plateau therefore generates a WI attractor at leading
order in which the dissipatively suppressed parameter $\eta_Q$ is
fixed entirely by $N_e$. (Our plateau potential is analogous in structure to the Starobinsky potential  \cite{starobinsky_1980, starobinskii_spectrum_1979} and $\alpha$-attractor
plateau potentials \cite{Kallosh_2013, ParticleDataGroup:2024cfk}.)  Since $\epsilon_{Q}=\mathcal{O}(N_{e}^{-2})$ while $\eta_{Q}=\mathcal{O}(N_{e}^{-1})$, the spectral tilt is dominated by
$\eta_{Q}$. For illustration, taking
$\{N_{e,\ast},\alpha,\Delta N_{\rm pre}\}=\{50,10,11.8\}$
gives $Q_\ast\simeq0.74$, and therefore
$|\eta_Q|\simeq2.0\times10^{-2}$ and
$\epsilon_Q\simeq3.49\times10^{-6}$. Note
that $\eta_{Q}<0$, as required for a red spectrum, because the plateau is concave. For the perturbation analysis we work in the temperature-independent sector of the WI slow-roll expansion. The inflaton potential [Eq.~\eqref{potentialscalar}] carries no explicit temperature dependence, hence the parameters of the general WI description vanish \cite{hall_moss_berera_2004, graham_moss_2009}: $b=T V_{,\chi T}/ V_{,\chi}=0,\,c=T\Upsilon_{,T}/\Upsilon=0$. Under these assumptions, the WI scalar-tilt expression reduces to the $b=c=0$ form. Within the leading order and near equilibrium WI perturbation treatment of Ref.~\cite{moss_xiong_2008}, specialized to $b=c=0$, the spectral index at arbitrary $Q$ is
\begin{align}
	n_{s} = 1 - \frac{9Q+17}{4(1+Q)}\epsilon_{Q}
	      - \frac{9Q+1}{4(1+Q)}\beta_{Q}
	      + \frac{3}{2}\eta_{Q} ,
	\label{spectralindex}
\end{align}
where $\beta_Q = (1+Q)^{-1} (d\ln\Upsilon/d\chi) (dV/d\chi)/{V}$ measures the variation of the dissipation coefficient along the background trajectory. From Eq.~\eqref{eq:UpsilonExact}, we obtain $d\ln\Upsilon/d\tau=-\gamma \Lambda_\infty /\Lambda$. Therefore, $\beta_Q$ does not vanish identically. However, using the slow-roll trajectory and $\kappa V\simeq3H^2$ in the observable regime, we obtain parametrically
\begin{align}
    \beta_Q &\simeq\frac{\gamma}{H}\frac{\Lambda_\infty}{\Lambda}
    \nonumber\\
    &=6Q\frac{\Lambda_\infty/\Lambda}{1-\Lambda_\infty/\Lambda}
    \simeq6Q\frac{\Lambda_\infty}{\Lambda}.
    \label{eq:betaSuppressed}
\end{align}
Since $\Lambda_\ast\gg\Lambda_\infty$, the horizon crossing value satisfies
$\beta_{Q,\ast}\ll1$, and we neglect it at the order considered below.
Eq.~\eqref{spectralindex} interpolates between the two dissipative regimes. Taking $Q\to\infty$ returns the following coefficients $9/4$, $9/4$ and $3/2$, reproducing the strong dissipative result of Ref.~\cite{hall_moss_berera_2004}, whose derivation can be found in that work, Eq.~(74), once the overall factor $1/r$ is absorbed into $\epsilon_{Q}=\epsilon_{V}/(1+Q)$. Taking $Q\to0$ returns $-\tfrac{17}{4}\epsilon_{V}+\tfrac32\eta_{V}-\tfrac14(d \ln \Upsilon/ d \chi) (dV/d\chi)/V$, which is the weak dissipative WI regime. Since our model sits at weak/strong crossover regime $Q\simeq1$, it is the interpolation rather than either limit which is doing the work here. Substituting Eq.~\eqref{eq:sreval} into Eq.~\eqref{spectralindex} gives the closed form result,
\begin{align}
    n_{s} \simeq
    1 - \frac{3}{2N_{e}}
      - \frac{9Q+17}{8\alpha^{2}N_{e}^{2}},
    \label{eq:nsclosed}
\end{align}
whose leading behavior is the attractor
\begin{align}
	n_{s} \simeq 1 - \frac{3}{2N_{e}}.
	\label{eq:nsattractor}
\end{align}
The leading term of Eq.~\eqref{eq:nsclosed} carries no dependence on $Q$, thus the attractor [Eq.~\eqref{eq:nsattractor}] holds throughout the WI regime. The coefficient $3/2$ is not new. A similar coefficient was obtained in the strongly dissipative regime for natural inflation \cite{visinelli_2011}. What is new in our model is that it survives on the plateau. We note first that Eq.~\eqref{potentialscalar} is an $\alpha$-attractor E-model, whose curvature parameter is $2/(3\alpha^{2})$ in our notation, hence our $\epsilon_Q$ and $\eta_Q$ [Eq.~\eqref{eq:sreval}] are similar with the E-model expressions \cite{Kallosh_2013}. In Ref.~\cite{chakraborty_das_2025}, the authors report that the attractor is lost for such plateau potentials in minimal WI. However, here the difference is quantitative. The correction term in Eq.~\eqref{eq:nsclosed} scales as $Q/\alpha^{2}$, and remains negligible only while $Q\ll\alpha^{2}N_{e}^{2}$. Our model enforces this through Eq.~\eqref{eq:Qpred}, fixing $Q\simeq1$ rather than $Q\gg1$, while $\alpha=\mathcal{O}(10)$ gives $\alpha^{2}N_{e}^{2}\simeq2\times10^{5}$. Moreover, in Ref.~\cite{chakraborty_das_2025}, the dissipation ratio is of order unity or larger, and the correction dominates the tilt.
\begin{table*}[!t]
\centering
\setlength{\tabcolsep}{10pt}
\begin{tabular}{@{}cccccccccc@{}}
\toprule
$N_{e,\ast}$ & $\Delta N_{\rm pre}$ & $Q_\ast$ & $\epsilon_Q$ &
$\eta_Q$ & $n_s$ & $r_{\mathcal T}^{\rm max}$ & $H_\ast$ &
$T_\ast/H_\ast$ & $r_{\mathcal T}$ \\
 & & & $[10^{-6}]$ & & & $[10^{-5}]$ &
$[10^{10}\,\mathrm{GeV}]$ & & $[10^{-8}]$ \\
\midrule
43 & 10.0 & 0.87 & 5.05 & $-0.0233$ & 0.9651 & 4.3 & 7.6 & 100 & 9.5 \\
50 & 11.8 & 0.74 & 3.49 & $-0.0200$ & 0.9700 & 3.2 & 7.0 &  94 & 7.9 \\
58 & 13.8 & 0.64 & 2.44 & $-0.0172$ & 0.9741 & 2.4 & 6.4 &  88 & 6.6 \\
60 & 14.3 & 0.62 & 2.25 & $-0.0167$ & 0.9750 & 2.2 & 6.2 &  86 & 6.3 \\
\bottomrule
\end{tabular}
\caption{WI slow-roll parameters $(\epsilon_Q,\eta_Q)$, spectral index $n_s$, inflationary Hubble scale at horizon crossing $H_\ast$, temperature-to-Hubble ratio $T_\ast/H_\ast$ and tensor-to-scalar ratio $r_{\mathcal T}$, in reduced Planck units with $\alpha=10$ and $g_\ast=106.75$. For each row, $\Delta N_{\rm pre}$ is chosen as the minimum value satisfying the source subdominance benchmark
Eq.~\eqref{eq:OmegaLambdaBenchmark}, corresponding to a suppression of
the direct vacuum decay source by at least two orders of magnitude.
The resulting $Q_\ast$ follows self-consistently from
$Q_\ast=46/(N_{e,\ast}+\Delta N_{\rm pre})$. The scale $H_\ast$ [Eq.~\eqref{eq:Hstar}] follows from the measured amplitude $A_s=2.1\times10^{-9}$ \cite{planck2018_VI}. The column $r_{\mathcal T}^{\rm max}=16\epsilon_Q/(1+Q_\ast)$ is the conservative bound obtained by setting $\mathcal F_\ast=1$ in Eq.~\eqref{eq:rtensor}, while $r_{\mathcal T}$ is the value obtained for $\mathcal F_\ast\simeq(3.5$--$4.6)\times10^{2}$. The row $N_{e,\ast}=58$ corresponds to the ACT DR6 determination of the tilt \cite{act_dr6_2025} and $N_{e,\ast}=43$ to Planck 2018 values \cite{planck2018_VI}.}
\label{TableEpsvEtav}
\end{table*}
The attractor of Eq.~\eqref{eq:nsattractor} should not be compared with
the cold inflation value $n_{s}\simeq1-2/N_{e}$, which the present model does not reproduce in any limit. Taking $Q\to0$ in Eq.~\eqref{spectralindex} gives $n_{s}=1-3/(2N_{e})-17/(8\alpha^{2}N_{e}^{2})$, since a thermal bath with $T>H$ modifies the spectrum even at vanishing dissipation.\\
\indent The measured value $n_s=0.974\pm0.003$ from the ACT DR6 power spectra \cite{act_dr6_2025} combined with Planck, CMB lensing and DESI DR1
baryon acoustic oscillations corresponds at leading attractor order 
to $N_{e,\ast}\simeq58$ within a $1\sigma$ interval
$N_{e,\ast}=52$--$65$. It is important, however, to distinguish this
observable number of e-folds from the total vacuum decay interval
$N_{e,\mathrm I}$ entering Eq.~\eqref{eq:Qpred}. Using
Eq.~\eqref{eq:NeDecomposition}, the dissipation ratio is
\begin{align}
    Q_\ast
    &\simeq \frac{46}{N_{e,\mathrm I}}
    =\frac{46}{N_{e,\ast}+\Delta N_{\rm pre}} .
    \label{eq:Qvalue}
\end{align}
Consequently, the spectral tilt determines $N_{e,\ast}$, whereas the
precise value of $Q_\ast$ depends on the duration of the initial
vacuum decay transient. For $\Delta N_{\rm pre}<N_{e,\ast}$, Eq.~\eqref{eq:Qvalue} may be expanded as
\begin{align}
    Q_\ast
    \simeq\frac{46}{N_{e,\ast}}\left[1-\frac{\Delta N_{\rm pre}}{N_{e,\ast}}+\mathcal O\!\left(\frac{\Delta N_{\rm pre}^{2}}{N_{e,\ast}^{2}}
    \right)\right].
    \label{eq:QtransientExpansion}
\end{align}
Thus, the central result $Q_\ast=\mathcal O(1)$ is insensitive to a
moderate pre-observable transient; nevertheless, its precise numerical
value is not fixed by $n_s$ alone. The tensor perturbations are not directly thermalized by the dissipative dynamics and retain their vacuum form, whereas the scalar curvature spectrum is enhanced by thermal and dissipative fluctuations \cite{benetti_ramos_2017}. We write the scalar enhancement factor at
horizon crossing as
\begin{align}
    \mathcal{F}_{\ast}
    = 1+2n_{\ast}  + \frac{2\sqrt{3}\pi Q_{\ast}}{\sqrt{3+4\pi Q_{\ast}}} \frac{T_{\ast}}{H_{\ast}},
    \label{eq:Fwarm}
\end{align}
where $ n_\ast = (\exp(H_\ast/T_\ast)-1)^{-1}$ denotes the inflaton Bose--Einstein occupation number near equilibrium. Since the dissipation coefficient adopted here has no explicit temperature dependence, $c=T\Upsilon_{,T}/\Upsilon=0$, we use the temperature-independent WI perturbation prescription, with the growing mode factor $G(Q_\ast)$ set to unity \cite{motaharfar_2019}. The enhanced growing mode associated with the coupled inflaton--radiation perturbations is relevant for temperature-dependent dissipation coefficients, $\Upsilon\propto T^c$ with $c\neq0$ \cite{graham_moss_2009,benetti_ramos_2017}. The use of this standard WI enhancement factor [Eq.~\eqref{eq:Fwarm}] assumes that during the observable window the thermal bath is dominated by the inflaton dissipative source rather than by direct vacuum decay. This requirement is quantified later by Eq.~\eqref{eq:OmegaLambdaStrong}. The tensor-to-scalar ratio is then
\begin{align}
    r_{\mathcal T}
    &= \frac{16\epsilon_V}
    {(1+Q)^2\mathcal F_\ast}
    =\frac{16\epsilon_Q}
    {(1+Q)\mathcal F_\ast}.
    \label{eq:rtensor}
\end{align}
Since the tensor spectrum remains proportional to $H^2$, its tilt is
controlled by the background expansion,
\begin{equation}
    \begin{split}
        n_{\mathcal T}
    &=\frac{d\ln\mathcal P_{\mathcal T}}{d\ln k}
    \simeq -2\epsilon_H \\
    &\simeq-2\left(\epsilon_Q+3Q\Omega_\Lambda
    \right),
    \end{split}
    \label{eq:ntensor}
\end{equation}
where Eq.~\eqref{eq:exitcondition} has been used. In the
potential-dominated observable regime, when
$3Q\Omega_\Lambda\ll\epsilon_Q$, this reduces to $n_{\mathcal T}\simeq-2\epsilon_Q$. Consequently, the cold single field consistency relation $r_{\mathcal T}=-8n_{\mathcal T}$ does not hold. In the observable regime where $\epsilon_H\simeq\epsilon_Q$, Eqs.~\eqref{eq:rtensor} and \eqref{eq:ntensor} instead give $r_{\mathcal T}\simeq-8n_{\mathcal T}/{(1+Q)\mathcal F_\ast}$, hence the scalar thermal enhancement suppresses $r_{\mathcal T}$ relative to $8|n_{\mathcal T}|$. Eq.~\eqref{eq:rtensor} shows that $r_{\mathcal T}$ is suppressed both by the plateau slow-roll parameter and by the dissipative factor $(1+Q)^{-1}$ multiplying $\epsilon_Q$ with an additional suppression from the scalar enhancement factor $\mathcal F_\ast$. Since $\mathcal F_\ast\geq1$, a model-independent conservative estimate of the tensor-to-scalar ratio within the present approximation is
\begin{align}
    r_{\mathcal T} \leq\frac{16\epsilon_Q}{1+Q}.
    \label{eq:rUpper}
\end{align}
A precise value requires the specification of the inflaton occupation
number $n_\ast$ and of the thermal ratio $T_\ast/H_\ast$, both of which are determined self-consistently in Sec.~\ref{subsec:exit}.
The resulting  estimates are summarized in Table~\ref{TableEpsvEtav}.

All (conservative) upper estimates given in Table~\ref{TableEpsvEtav}
lie nearly three orders of magnitude below the current bound
$r_{\mathcal T}<0.036$
\cite{Ade_2021,planck2018_X,ParticleDataGroup:2024cfk}. Since
$\mathcal F_\ast\geq1$, the physical tensor-to-scalar ratio is further
suppressed relative to these values. The model therefore predicts a
tensor signal below current sensitivity within the perturbative
prescription adopted here. The spectral index selects the observable slow-roll interval $N_{e,\ast}$: $N_{e,\ast}\simeq58$ for the ACT determination and $N_{e,\ast}\simeq43$ for Planck 2018 at leading attractor order. The hierarchy, however, constrains the combination
$Q_\ast(N_{e,\ast}+\Delta N_{\rm pre})\simeq46$. Consequently, $Q_\ast\simeq1$ for a moderate pre-observable transient, while its precise value highly depends strongly on $\Delta N_{\rm pre}$ through Eq.~\eqref{eq:dNpremin}. If, rather, a lower value were preferred by future data, it would remain natural within this model rather than problematic, since the vacuum energy drains into the radiation bath throughout inflation [Eq.~\eqref{MRContEquations}]. The bath is thus already populated at exit, reducing the post-inflationary expansion required to reach the observed CMB scales. In either situation, the determination is falsifiable and it demands a full numerical solution of the FL equations and WI dynamics. Next, we discuss quantitatively the graceful exit from inflation.
\subsection{The graceful exit}
\label{subsec:exit}
\noindent The first Hubble slow-roll parameter is defined as
\begin{align}
	\epsilon_{H}
	= -\frac{\dot H}{NH^{2}} =-\frac{1}{H^{2}}\frac{dH}{d\tau},
	\label{eq:epsilonH}
\end{align}
the factor $1/N$ arising for the same reason as in Eq.~\eqref{q0dissipation}: the derivative must be taken with respect to proper time for $\epsilon_H=1$ to mark the end of inflation. Using the quasi-stationary solution of Eq.~\eqref{MRContEquations},
$e^{-\phi/2}\rho_{\rm rad}\simeq\tfrac34 Q\chi'^{2}+\tfrac32 Q\rho_{\Lambda}$~\cite{footnote1}, together with the slow-roll relation for $\chi'$, Eq.~\eqref{eq:epsilonH} gives
\begin{align}
	\epsilon_{H} \simeq \epsilon_{Q} + 3Q\Omega_{\Lambda} ,
	\quad
	\Omega_{\Lambda} = \frac{\Lambda}{3H^{2}} .
	\label{eq:exitcondition}
\end{align}
Eq.~\eqref{eq:exitcondition} is therefore an analytic
slow-roll estimate of the exit condition rather than an exact background identity. Inflation proceeds while $\epsilon_{H}<1$ and ends at $\epsilon_{H}\simeq1$. Two regimes are worth discussing: (I) for $3Q\Omega_{\Lambda}\ll1$ the dynamics is governed by the potential and inflation terminates when the inflaton leaves the plateau, $\epsilon_{Q}\to1$; (II) if instead the vacuum energy is not subdominant, $\Omega_{\Lambda}\to1$ drives the exit, in which case Eq.~\eqref{eq:exitcondition} reduces to $\epsilon_{H}=3Q$ and inflation is impossible for $Q>1/3$.
Consistency therefore requires
\begin{align}
	\Omega_{\Lambda} \ll \frac{1}{3Q} \simeq \frac{1}{3} ,
	\label{eq:OmegaLambdacondition}
\end{align}
at the predicted weak/strong crossover regime $Q\simeq1$. Eq.~\eqref{eq:OmegaLambdacondition} is not in tension with the premise of our model, and it is worth discussing why. The dissipation ratio depends on the logarithmic decay rate [Eq.~\eqref{q0dissipation}], which is scale free, whereas $\Omega_{\Lambda}$ measures the vacuum contribution of the energy budget. The two are independent. A vacuum component that is dynamically negligible may nonetheless decay rapidly in relative terms and thereby supply substantial friction. Under Eq.~\eqref{eq:OmegaLambdacondition}, the graceful exit is governed by the potential, and the vacuum energy continues to decay towards $\Lambda_{\infty}$ throughout inflation, feeding the radiation bath [Eq.~\eqref{MRContEquations}]. For the perturbation analysis of Sec.~\ref{subsec:observables}, a stronger
condition is required during the observable slow-roll window. The
conformally weighted radiation density contains two source terms, $e^{-\phi/2}\rho_{\rm rad}\simeq \frac{3}{4}Q\chi'^2+\frac{3}{2}Q\rho_\Lambda$, corresponding respectively to inflaton dissipation and direct vacuum decay. Without a vacuum decay channel, conventional effective warm inflation models only retain the first term, $\frac{3}{4} Q \chi^{\prime 2}$. Our model, however, accounts for an additional contribution to the radiation bath density stemming from vacuum decay. This approach is based on an energy balance argument within a phenomenological model, rather than a microscopic derivation of the thermal bath. Furthermore, a near equilibrium thermal bath requires rapid interaction rates in addition to the condition $T > H$, which is consistently satisfied in our calculations [see Table. 1]. We want to specify that the exact particle content, interaction rates, and fluctuation-dissipation completion are not specified. Since the standard WI perturbation spectrum employed above assumes that the thermal bath is predominantly sustained by the dissipative
inflaton source, we impose at horizon crossing $ \left|\rho_\Lambda'\right|\ll \Upsilon\chi'^2$. Using $\rho_\Lambda'=-6HQ\rho_\Lambda$ and $\Upsilon=3HQ$, this condition becomes $\rho_\Lambda\ll\frac{1}{2}\chi'^2$. With
$\chi'^2\simeq2\epsilon_QH^2(1+Q)^{-1}$, it can equivalently be written as
\begin{align}
    \Omega_{\Lambda,\ast} \ll \frac{\epsilon_{Q,\ast}}
    {3(1+Q_\ast)}.
    \label{eq:OmegaLambdaStrong}
\end{align}
We stress that Eq.~\eqref{eq:OmegaLambdacondition} is only the condition for accelerated expansion. The stronger condition given in Eq.~\eqref{eq:OmegaLambdaStrong} is imposed specifically during the
observable window in order for the radiation bath and its perturbations
to be described by the WI treatment adopted in Sec.~\ref{subsec:observables}. \\
\indent To quantify the source subdominance requirement, we explicitly introduce a dimensionless tolerance $\delta\ll1$ through $ |\rho_{\Lambda,\ast}'|/(\Upsilon_\ast\chi_\ast'^2)\leq \delta$. Using $|\rho_\Lambda'|=6HQ\rho_\Lambda$, $\Upsilon=3HQ$, and $\chi'^2\simeq2\epsilon_QH^2(1+Q)^{-1}$, this becomes
\begin{align}
    \Omega_{\Lambda,\ast}
    \leq
    \delta\,
    \frac{\epsilon_{Q,\ast}}
    {3(1+Q_\ast)} .
    \label{eq:OmegaLambdaBenchmark}
\end{align}
For the numerical benchmarks in Table~\ref{TableEpsvEtav} we adopt
$\delta=10^{-2}$, where the factor $\delta^{-1}=10^{2}$ requires the
direct vacuum decay source to contribute at most one percent of the
inflaton dissipative source. Under the source subdominance conditions
Eqs.~\eqref{eq:OmegaLambdaStrong} and \eqref{eq:OmegaLambdaBenchmark},
the direct vacuum decay contribution to the radiation bath is
subleading relative to the inflaton dissipation term, hence
$e^{-\phi/2}\rho_{\rm rad}\simeq 3Q\chi'^{2}/4$.
Consistently with Eq.~\eqref{eq:effdens}, the thermal radiation
density entering the Einstein frame description is the conformally
weighted combination $e^{-\phi/2}\rho_{\rm rad}=(\pi^{2}/30)g_\ast T^{4}$. Using $\chi'^{2}\simeq2\epsilon_QH^{2}(1+Q)^{-1}$ then gives the WI condition
\begin{align}
    \frac{T}{H}= \left[\frac{45 Q \epsilon_Q}{\pi^{2}g_\ast(1+Q)}\right]^{1/4}H^{-1/2},
    \label{eq:warmcondition}
\end{align}
in reduced Planck units. It remains to fix the inflationary scale and to verify the WI condition $T>H$ [Eq.~\eqref{eq:warmcondition}]. The observed scalar amplitude fixes the normalization of the inflationary scale. Using
$\mathcal P_\zeta= (H^{2}/(2\pi\chi'))^2\mathcal F_\ast$ together with
$\chi'^{2}\simeq 2\epsilon_QH^{2}(1+Q)^{-1}$, one obtains, in reduced Planck units,
\begin{align}
    H_\ast^{2}
    = \frac{8\pi^{2}\epsilon_{Q,\ast}A_s}
    {(1+Q_\ast)\mathcal F_\ast},
    \label{eq:Hstar}
\end{align}
where $A_s=2.1\times10^{-9}$ \cite{planck2018_VI}. Since $\mathcal F_\ast\geq1$, Eq.~\eqref{eq:Hstar} immediately provides an upper bound on the inflationary scale. For example, for the $N_{e,\ast}=43$ row of Table~\ref{TableEpsvEtav}, where $Q_\ast\simeq0.87$ and $\epsilon_{Q,\ast}\simeq5.05\times10^{-6}$, one obtains $H_\ast \lesssim1.6\times10^{12}\,\mathrm{GeV}$. To determine the physical value of $H_\ast$ together with the thermal enhancement, we assume a near-equilibrium Bose--Einstein occupation number for the inflaton. We then solve Eqs.~\eqref{eq:Fwarm}, \eqref{eq:warmcondition}, and \eqref{eq:Hstar} with $g_\ast=106.75$.
For the parameter range displayed in Table~\ref{TableEpsvEtav} this
gives
\begin{equation}
    \begin{split}
    \mathcal F_\ast&\simeq(3.5\text{--}4.6)\times10^{2},
    \\
    H_\ast&\simeq(6.2\text{--}7.6)\times10^{10}\,\mathrm{GeV},
    \\
    \frac{T_\ast}{H_\ast}&\simeq86\text{--}100 .
    \end{split}
\end{equation}
Thus the WI condition $T_\ast>H_\ast$ is comfortably satisfied over
the representative observable range. The scalar amplitude normalization allows us to derive a lower bound on $\Delta N_{\rm pre}$. From Eq.~\eqref{eq:LambdaNe}, the vacuum contribution at
horizon crossing is $\Lambda_\ast =\Lambda_{\rm I}e^{-6Q_\ast\Delta N_{\rm pre}}$, since $N_{e,\rm I}-N_{e,\ast}=\Delta N_{\rm pre}$.
To ensure that the standard WI perturbation treatment is safely
applicable, we impose the quantitative source subdominance condition
Eq.~\eqref{eq:OmegaLambdaBenchmark}. Using
$\Omega_{\Lambda,\ast}=\Lambda_\ast/(3H_\ast^2)$
together with Eq.~\eqref{eq:Hstar}, one obtains
\begin{align}
    \Delta N_{\rm pre}
    \geq \frac{1}{6Q_\ast} \ln\!\left[\frac{\Lambda_{\rm I}(1+Q_\ast)^2\mathcal F_\ast}{8\pi^2\delta\,\epsilon_{Q,\ast}^2A_s
    }\right].
    \label{eq:dNpremin}
\end{align}
Solving the consistency condition Eq.~\eqref{eq:dNpremin}, the ACT value $N_{e,\ast}\simeq58$ gives
$\Delta N_{\rm pre}\simeq13.8$ and hence, from Eq.~\eqref{eq:Qvalue}, $Q_\ast\simeq0.64$. Likewise, the Planck 2018 value $N_{e,\ast}\simeq43$ gives $\Delta N_{\rm pre}\simeq10.0$ and
$Q_\ast\simeq0.87$. Since $Q_\ast=46/(N_{e,\ast}+\Delta N_{\rm pre})$ and $\mathcal F_\ast$ depends on $Q_\ast$, then Eq.~\eqref{eq:dNpremin} is an implicit condition, which is solved for each $N_{e,\ast}$. The values displayed in Table~\ref{TableEpsvEtav} are the minimum transient durations satisfying the condition given in Eq.~\eqref{eq:OmegaLambdaBenchmark}. The actual duration will be determined by the complete background evolution, which can only lengthen the transient duration and lower $Q_\ast$. These values position our model in the weak/strong dissipative crossover. For the numerical estimates, we normalize the primordial vacuum scale to $\Lambda_{\rm I}=1$ in reduced Planck units, corresponding to $\Lambda_\infty\simeq10^{-120}$ for the hierarchy adopted in Eq.~\eqref{eq:Qpred}. The dependence on $\mathcal F_\ast$ is only logarithmic: changing $\mathcal F_\ast$ by one decade changes the lower bound by approximately $0.4$--$0.6$ e-folds over the range of $Q_\ast$ considered here. A complete determination of $\{\rho_{\rm rad}/\rho_{\chi}(N_e),T(N_e),H(N_e)\}$ requires the full numerical treatment, which is deferred to the second part of this work.
\section{Discussion}\label{sec:summary}
The WIS framework developed in Sec.~\ref{EFEWeyl} provides a geometric
route to the early- and late time dynamics of the Universe. The scalar
field $\phi$ is the potential of
the non-metricity  $w_\mu=\partial_\mu\phi$, and is therefore a
natural geometric degree of freedom. We emphasize, however, a point that
is easily overlooked. Commonly in a Weyl invariant theory, $\phi$ is a
compensator and can be gauged away. However, here it acquires physical content
only because the volume element $\sqrt{-g}$ breaks the symmetry, leaving
the residual kinetic normalization $\lambda=3/2-2\xi$ [Eq.~\eqref{eq:action_einstein}]. For $\lambda=0$, i.e. $\xi=3/4$, the kinetic degree of freedom of the scalar disappears, and the canonical inflaton construction used in our model ceases to apply.
The inflaton (a redefinition of $\phi$ [cf. Eq.~\eqref{eq:KGproper}]) of our model is thus geometric in origin but physical only by virtue of broken Weyl invariance, and we restricted this work to $\lambda>0$ for quintessence models. The origin of that breaking is the measure: under Eq.~\eqref{eq:weylgauge} one has
$\sqrt{-g}\to e^{2\Omega}\sqrt{-g}$ while $\tilde{R}=g^{\mu\nu}\tilde{R}_{\mu\nu}\to
e^{-\Omega}\tilde{R}$, hence $\sqrt{-g}\tilde{R}\to e^{\Omega}\sqrt{-g}\tilde{R}$.
The invariant measure would be $\sqrt{-\tilde{g}}$; our choice of $\sqrt{-g}$ in
Eq.~\eqref{eq:action_weyl} fixes the Weyl frame, and the breaking is therefore explicit
rather than spontaneous.\\ 
\indent The distinctive feature of our model is that the
dynamical vacuum is used both as a source in the EFEs and
as the quantity defining the effective WI dissipation coefficient.
Within the lapse gauge of Eq.~\eqref{eq:scaleinvariant}, its prescribed
history is also encoded in the time parametrization through
$N\propto\sqrt{\Lambda}$. We stress that the lapse itself is not an
additional physical degree of freedom and all physical statements are made
in terms of proper time quantities.\\
\indent A dissipation ratio [cf. Eq.~\eqref{q0dissipation}] tied to the
dynamical vacuum sector is identified with the corresponding quantity of
the WI framework \cite{berera_warm_1995, bastero_gil_2021}. As a consequence, the following results emerge: (I) the WI dissipation coefficient is identified with one half of the logarithmic decay rate of the vacuum energy [cf. Eq.~\eqref{qdissipation}]. In the standard WI constructions the dissipation coefficient $\Upsilon(T)$ encodes the microphysics of inflaton--bath interactions \cite{berera_moss_ramos_2009, bastero_gil_2016, bastero_gil_2021} while in our setup it is phenomenologically identified with the logarithmic decay rate of the dynamical vacuum. Within our dynamical vacuum hypothesis [Eq.~\eqref{eq:Lambdaansatz}], one finds $\Upsilon\simeq\gamma/2$ approximately constant over the observable interval, placing this model in the constant dissipation class \cite{bastero_gil_2021}; (II) The choice of $V(\phi)$ with a plateau [Eq.~\eqref{potentialscalar}] yields the attractor 
$\eta_Q=-1/N_e$, whence $n_s\simeq1-3/(2N_e)$ [Eq.~\eqref{eq:nsattractor}]. The tensor-to-scalar ratio is suppressed by three
independent factors and lies orders of magnitude below current bounds
[Eq.~\eqref{eq:rtensor}]; (III) The dissipation ratio is itself fixed by the cosmological constant hierarchy through $QN_{e,\mathrm{I}}\simeq46$ [Eq.~\eqref{eq:Qpred}]. 
The hierarchy therefore does not allow $Q$ to be scanned independently: it constrains the combination $Q_\ast\left(N_{e,\ast}+\Delta N_{\rm pre}\right)\simeq46$. The spectral tilt determines $N_{e,\ast}$, while the transient duration $\Delta N_{\rm pre}$ is bounded from below by
Eq.~\eqref{eq:dNpremin}, which depends on the perturbative
normalisation $\mathcal F_\ast$ only logarithmically. Our model
therefore selects $Q_\ast\simeq0.6$--$0.9$ across the observationally
allowed range of $N_{e,\ast}$, the residual freedom being fixed by the
full background integration. 

\indent A closer antecedent deserves explicit mention. The SIV hypothesis \cite{Maeder2017, maeder_scale-invariant_2020, universe11020048, maeder_gueorguiev_2021} is likewise built on Weyl integrable geometry, treats $\Lambda$ as a consequence of the time parametrization rather than as an independent constant, and has already been applied to inflation with a graceful exit \cite{maeder_gueorguiev_2021}. Our model departs from theirs in the dissipative sector. In their work, the vacuum sets the expansion but no radiation bath is generated and no friction beyond the Hubble term appears, whereas in our work the logarithmic evolution of the same lapse that carries $\Lambda$ is identified with the effective WI dissipation coefficient through Eq.~\eqref{qdissipation}, from which the integrated constraint $Q N_{e,\rm I}\simeq46$ under the approximately constant $Q$ and the observables of Sec.~\ref{subsec:observables} emerge. In other words, in the SIV construction the dynamical vacuum affects the background evolution without introducing a WI bath, whereas in the present WI effective embedding its logarithmic decay rate is additionally identified with a dissipative coefficient that sources the radiation sector.\\
\indent It is worth contrasting the two-de Sitter architecture of Eq.~\eqref{eq:Lambdaansatz} with the running vacuum models $\Lambda=\Lambda(H)$ of Ref.~\cite{PhysRevD.88.063531}, which likewise interpolate between a primeval and a present de Sitter phase and obtain a graceful exit. The mechanisms differ in an essential respect. Their dynamic $\Lambda$ is a function of the Hubble rate: $\Lambda=c_0+c_2H^2+c_4H^4/H_{\textrm{I}}^2$ and the decay is a dynamical consequence of the expansion itself. In our work, the dynamic $\Lambda$ is tied to the lapse through Eq.~\eqref{eq:scaleinvariant}, thereby yielding a logarithmic decay rate that is further identified with the dissipation coefficient [Eq.~\eqref{qdissipation}]. This identification has no counterpart in
Ref.~\cite{PhysRevD.88.063531}. Combined with the assumed vacuum
hierarchy and the approximately constant-$Q$ decay interval, it leads to
the constraint Eq.~\eqref{eq:Qpred}.\\
\indent A further point of contact deserves mention. The effective WI description is widely argued to evade the de Sitter swampland conjecture, or strengthen the de Sitter constraint \cite{Brandenberger_2020}, since the dissipative friction relaxes the steepness required of the potential \cite{motaharfar_2019, das_2019, bedroya_vafa_2020}. We note in this connection that a vacuum energy which relaxes over the inflationary interval is more readily accommodated within the swampland
constraints than a stabilized de Sitter vacuum.
In our model, the effective WI dissipation is not derived from a microscopic inflaton--bath interaction but instead, its coefficient is identified with the logarithmic decay rate of the dynamical vacuum [Eq.~\eqref{qdissipation}]. Consequently, once the vacuum hierarchy and the decay interval are specified, the associated dissipation ratio is
constrained to $Q=\mathcal O(1)$ rather than freely scanned.\\
\indent The inflationary calibration
$\gamma\simeq6QH_{\rm I}$ implies $\gamma/H_0\gg1$. The late time constancy of the vacuum energy instead follows from the exponential suppression of the transient:
for $\gamma\tau_0\gg1$ one has $\Lambda(\tau_0)\simeq\Lambda_\infty$, yielding $d\ln\Lambda/d\tau\simeq-\gamma(\Lambda-\Lambda_\infty)/\Lambda\rightarrow0$. Hence, the vacuum becomes observationally indistinguishable from a cosmological constant at low redshift even though the microscopic decay scale $\gamma$ is very large.  Equivalently, Eq.~\eqref{eq:UpsilonExact} gives $\Upsilon\rightarrow0$ as $\Lambda\rightarrow\Lambda_\infty$, therefore the effective WI dissipation naturally switches off after the primordial vacuum decay epoch. This effectively gives a de Sitter architecture of the model. The authors in Ref.~\cite{Brandenberger_2020}, perform an extensive study for single scalar field theories in the WI description and considering the entropy production they reach to the conclusion that the de Sitter constraint is safely obtained. \\
\indent It is worth commenting that $\phi$, being the potential of the non-metricity and coupling universally to matter under the weight $e^{-\phi/2}$ [Eq.~\eqref{eq:Tmunu}] with strength $\beta=(2\sqrt{\lambda})^{-1}$ in Planck units, shares the defining features of a dilaton. The conformal relation $\tilde g_{\mu\nu}=e^{\phi}g_{\mu\nu}$ is of the same form as that connecting the string and Einstein frames, and the universality of the matter coupling is likewise the characteristic property which renders a light dilaton subject to fifth-force and equivalence principle constraints. In our model, however, Eq.~\eqref{potentialscalar} gives the scalar mass $m_\chi^{2}=2\alpha^{2}V_0$ at its minimum, hence $\chi$ is massive and short-ranged after inflation, and the late time acceleration is carried by $\Lambda_\infty$ rather than by $\phi$. Whether the dilatonic analogy extends beyond this structural similarity, and in particular whether the relaxation of $\Lambda$ studied here admits a string theoretic interpretation, will be studied in future work.\\
\indent The Weyl structure underlying our model has a suggestive counterpart in holographic renormalization. In the Weyl--Fefferman--Graham formulation \cite{ciambelli_leigh_2020,jia_karydas_2021}, the boundary geometry inherits a Weyl connection rather than only a Levi--Civita connection, while radial bulk evolution provides the geometric realization of renormalization group (RG) flow \cite{deboer_2000}. We stress, however, that Eq.~\eqref{eq:scaleinvariant} has not been derived holographically; its interpretation as an RG trajectory is an analogy to be tested in a complete dual formulation. A formal background-level dictionary can be constructed. Following the usual holographic interpretation of cosmological evolution, we introduce the following: $\beta_\chi=d\chi/d\ln\mu$, with $\ln\mu\leftrightarrow-N_e$, where $\mu$ denotes the boundary RG scale. Since $dN_e=-H d\tau$, this convention gives $\beta_\chi =-d\chi/dN_e =\chi'/H$. In a cold single field inflationary system one has $\epsilon_H=\beta_\chi^2/2$. This identity is modified in the WI background used here, because both dissipation and the dynamical vacuum contribute to the Hubble evolution. Using Eqs.~\eqref{eq:exitcondition} and the slow-roll approximation
$\chi'^2\simeq 2\epsilon_QH^{2}(1+Q)^{-1}$ gives
\begin{align}
    \epsilon_H\simeq \frac{1+Q}{2}\beta_\chi^2
    + 3Q\Omega_\Lambda .
    \label{eq:holoBetaWarm}
\end{align}
Therefore, the dissipation ratio is not identified directly with
$\beta_\chi^2$. Instead, $Q$ modifies the relation between the scalar
RG flow and the Hubble slow-roll parameter.
In the observable source subdominant regime,
Eq.~\eqref{eq:OmegaLambdaStrong} implies
$3Q\Omega_\Lambda\ll\epsilon_Q$, so that
Eq.~\eqref{eq:holoBetaWarm} reduces to $\epsilon_H\simeq(1+Q)\beta^{2}_{\chi}/2\simeq\epsilon_Q$. Moreover, the plateau trajectory of Eq.~\eqref{eq:trajectory} gives $\beta_\chi=-d\chi/dN_e \simeq-1/(\alpha N_e)$, thus giving $\beta_\chi^2\simeq 1/(\alpha N_e)^2= 2\epsilon_Q(1+Q)^{-1}$. The scalar RG flow is consequently close to a fixed point during the observable plateau, while the finite value $Q=\mathcal O(1)$ controls the dissipative modification of the flow rather than the beta function itself.
\\
\indent The Weyl group structure [Sec.~\ref{subsec:weyl}] admits a
formal RG interpretation. Local Weyl transformations of the
boundary metric provide the geometrical language of local
RG transformations \cite{osborn_1991}. Thus, we regard $\mathrm{Weyl}(\mathcal M)$ as a geometric analogue of the local RG transformations rather than asserting a literal duality of groups. Bulk integrability corresponds to a closed Weyl connection, locally pure gauge and therefore of trivial scale holonomy \cite{ciambelli_leigh_2020,jia_karydas_2021}. Because $N$ is a lapse-gauge quantity, the following construction is understood only after the gauge condition [Eq.~\eqref{eq:scaleinvariant}] is fixed. In the gauge $N(t_0)=1$, Eq.~\eqref{eq:scaleinvariant} permits the
formal parametrization $\Lambda/\Lambda(t_0)=e^\Omega$. From Eq.~\eqref{eq:QNe}, we obtain $d\Omega/dN_e=d\Omega/d\ln\mu=-6Q$ where the convention $\ln\mu\leftrightarrow-N_e$ is again used. For an approximately constant $Q$,
\begin{align}
    \Lambda(\mu)
    \simeq
    \Lambda_\ast
    \left(
    \frac{\mu}{\mu_\ast}
    \right)^{-6Q}.
    \label{eq:LambdaRG}
\end{align}
Hence, $Q$ measures the logarithmic rate of the vacuum trajectory along
the Weyl/RG flow. The late time limit is defined dynamically by $d\ln\Lambda/d\tau\rightarrow0$ and $\Upsilon\rightarrow0$, which corresponds formally to a fixed point of the vacuum flow. Whether this fixed point is to be assigned a UV or IR interpretation highly depends on the domain-wall/cosmology continuation and is not fixed by the background evolution. We regard this last identification as an interpretation rather than a derivation, since a dual operator for $\Lambda$ requires the promotion $\Lambda\to\Lambda(\phi)$ permitted by Eq.~\eqref{eq:Lambdaeff}.\\ 
\indent A direct application of AdS/CFT \cite{maldacena_1998} is not available, since in our model we assumed $\Lambda>0$ and the spacetime is asymptotically de Sitter. However, three adjacent frameworks are applicable: 1) The domain-wall/cosmology correspondence \cite{skenderis_townsend_2006} maps FLRW backgrounds with exponential-type potentials onto Euclidean AdS domain walls, for which the dual descriptions are known \cite{kanitscheider_2008}; 2) holographic cosmology
\cite{mcfadden_skenderis_2010} computes $n_s$ and $r_{\mathcal{T}}$ directly from boundary correlators of the stress tensor, and is particularly well matched to the present setting because a four-dimensional bulk is associated with a three-dimensional boundary theory, for which the standard local Weyl anomaly of even-dimensional CFTs is absent; 3) dS/CFT \cite{strominger_2001} applies to the asymptotic future, where $\Lambda(t)\to\Lambda_{\infty}$ defines an exact de Sitter fixed point. The principal obstruction is that $\Lambda(t)$ is prescribed externally and has no dual operator. Promoting $\Lambda\to\Lambda(\phi)$, as permitted by the degeneracy Eq.~\eqref{eq:Lambdaeff}, would remove it. We regard holographic cosmology as the most promising of the three, and it will be explored in future work. We also note that the same promotion would be required of any covariant quantum treatment: our framework is used as an effective field theory below $M_{\rm Pl}$, and its RG properties are not addressed here. \\
\indent As an outlook, several quantities identified analytically here require numerical determination. The full integration of Eqs.~\eqref{FL1}--\eqref{MRContEquations} together with Eq.~\eqref{eq:KGproper} will fix the number of e-folds $N_e$, the horizon crossing epoch, and thereby convert the attractor [cf. Eq.~\eqref{eq:nsattractor}] into a definite prediction. The bath temperature must be computed from the quasi-stationary radiation density to confirm the estimate of Eq.~\eqref{eq:warmcondition} and the WI condition
$T> H$ throughout the trajectory. The late time trajectory is needed to determine $\phi(t_0)$. Finally, the
inclusion of a matter component, the calculation of the B-mode
polarisation of the CMB, and the holographic reconstruction already outlined are natural extensions. Numerical work is not expected to alter the qualitative mechanisms identified here: the decoupling of $Q$ from $\epsilon_H$, the dissipation coefficient as the logarithmic decay rate of the dynamic vacuum energy [Eq.~\eqref{qdissipation}], the constraint $QN_{e,\mathrm I}\simeq46$ and $n_s\simeq1-3/(2N_e)$, but is required for quantitative precision and predictive power.
\section{Conclusions}\label{conclusions}
We have investigated a dynamic vacuum scenario within the WIS framework. The SIV hypothesis [Eq.~\eqref{eq:scaleinvariant}] is applied together with the ansatz for $\Lambda(t)$ [Eq.~\eqref{eq:Lambdaansatz}], thus linking spacetime geometry, FL cosmology and the effective WI dynamics. Within WIS, the scalar field $\phi$ enters as the potential of the non-metricity function and is therefore geometric in origin. In addition, it becomes physical only through the breaking of Weyl invariance. \\ \indent Five main results then follow from this framework: (I) An exchange between
the vacuum and non-vacuum sectors is forced by $\nabla^{\mu}G_{\mu\nu}=0$, and the resulting cancellation of the explicit source $\dot\rho_\Lambda$ in the acceleration equation leaves the Hubble slow-roll parameter controlled solely by the non-vacuum content, while $\Lambda$ itself survives in the observables through the expansion rate. A dissipation ratio is introduced [Eq.~\eqref{q0dissipation}], resembling the dynamics of warm inflation. 
(II) We identify the WI dissipation coefficient with
one half of the logarithmic decay rate of the vacuum energy
[Eq.~\eqref{qdissipation}]. This quantity is approximately constant over
the observable interval for the vacuum law adopted here.
(III) The concave-shaped inflaton potential [Eq.~\eqref{potentialscalar}] generates the attractor $\eta_Q=-1/N_e$, yielding $n_s\simeq1-3/(2N_e)$, for which the slope of the potential and the dissipation dependence cancel at leading order. The tensor-to-scalar ratio is several orders of magnitude below current bounds [Eqs.~\eqref{eq:sreval}--\eqref{eq:rtensor}]. (IV) The dissipation ratio is not an independently scanned parameter: the total decay of $\Lambda$ fixes $Q N_{e,\mathrm I}\simeq46$, with $N_{e,\mathrm I}=N_{e,\ast}+\Delta N_{\rm pre}$ [Eqs.~\eqref{eq:Qpred} and \eqref{eq:NeDecomposition}]. The spectral tilt fixes the observable interval $N_{e,\ast}$, whereas the pre-observable transient $\Delta N_{\rm pre}$ is bounded from below by the source subdominance condition
Eq.~\eqref{eq:dNpremin}. Consequently, our model robustly selects $Q_\ast\simeq0.6$--$0.9$, with the full background evolution fixing the residual freedom. 

(V) Within the slow-roll approximation, the graceful exit admits the analytic estimate
$\epsilon_H\simeq\epsilon_Q+3Q\Omega_\Lambda$ where inflation ends when the exact condition $\epsilon_H=1$ is reached [Eq.~\eqref{eq:exitcondition}].\\
\indent The adopted vacuum law [Eq.~\eqref{eq:Lambdaansatz}] contains a rapidly decaying primordial component and a nonzero asymptotic remainder $\Lambda_\infty$. Since the inflationary calibration gives $\gamma\sim H_{\rm I}\gg H_0$, the transient is exponentially suppressed by the present epoch, $\gamma\tau_0\gg1$, leaving $\Lambda(t_0)\simeq\Lambda_\infty$ and $\Upsilon(t_0)\simeq0$. The Universe therefore approaches a de Sitter fixed point with vanishing vacuum--radiation exchange.

\indent This asymptotic
regime is also where the model makes contact with gauge/gravity duality:
the integrable Weyl structure is the bulk counterpart of a boundary Weyl
connection, while the scalar trajectory admits a formal holographic
beta-function description. In the WI background the usual cold
single field relation is modified to
$\epsilon_H\simeq(1+Q)\beta_\chi^2/2+3Q\Omega_\Lambda$
[Eq.~\eqref{eq:holoBetaWarm}], so that $Q$ modifies the RG flow rather
than being directly identified with the squared beta function.

Our model provides an analytically motivated parameter window for the cosmological observables. A quantitatively complete prediction requires the numerical determination of the early time trajectory, the end of inflation and the horizon crossing epoch, together with verification of the WI condition $T>H$ and of the late time constraint on the matter coupling. These will be addressed in the second part of this work.
\bibliography{sn-bibliography}
\end{document}